\documentclass[11pt]{article}
\usepackage{fleqn,cospar}

\title{COSMOLOGICAL MODELS: THEORY AND OBSERVATIONS}

\author{V.N.Lukash\address{Astro Space Center of Lebedev Physical
Institute (ASC FIAN),\\ 
    84/32 Profsoyuznaya st, Moscow, 117810, Russia}}
\begin{document}
				    
\maketitle
				    
\begin{abstract}
The current status of the theoretical and observational cosmology 
is reviewed.
\end{abstract}

\section*{INTRODUCTION}

Investigation of the anisotropy of the relic CMB radiation ensures a basic 
channel of the information about the early Universe. The continuing 
progress in the technology of deep-sky galactic surveys has resulted 
in impressive knowledge on the {\em Large Scale Structure} in the Universe 
as well as its evolution back to high redshifts. Both $\Delta T/T$ and 
LSS observational data complement on supercluster scales and thus disclose 
the cosmological model. Geometrical experiments (SNIa, Deuterium) provide 
additional independent tests hinting on a complex nature of the dark matter 
in the Universe.

The cosmological model gets parameters. There is a persistent evidence that 
\begin{itemize}
\item
the underlying space is flat,
\item
the Hubble parameter is close to $65\;{\rm km}\;{\rm s}^{-1} {\rm Mpc}^{-1}$ 
which is consistent with the age of old globular clusters if $\Omega_m\le 0.5$,
\item
the structure in the Universe has developed in the process of gravitational
instability from small Gaussian perturbations of matter density with a scale 
invariant fundamental spectrum.
\end{itemize}

The situation in cosmology is currently dominated by observations,
however the accuracy of present observational data is still too low to 
determine a unique set of all cosmological parameters. The main physical issue
--- a nature and composition of dark matter --- is still under discussion. 
A great hope of cosmologists is related with the development of ground 
and space based $\Delta T/T$ and far Universe observations which will 
help to delimit the cosmological model up to a few per cent in the nearest 
future.

\section*{FORMATION OF LSS IN THE UNIVERSE}

The seeds of the visible large scale structure in the Universe are the 
{\em Cosmological Density Perturbations} which grow due to gravitational 
instability in the late period of the Universe expasion when density is 
dominated by cold non-relativistic matter ($z_\sim^<\,10^4$). These primordial 
CDPs must have been created at the inflationary Big Bang epoch, as after 
the end of inflation the Universe was hot (radiationally dominated) 
and thus gravitationally stable against small perturbations of 
matter density and gravitaty field. The CDPs existed in the hot Universe 
as longwave 'gravitating sound waves' propagating across the relativistic 
matter with a constant amplitude. 

The CDPs started growing only after the equality epoch when the matter 
pressure decayed. The required fundamental CDP for galaxy clusters to be 
formed by now, has therefore been predicted at the level $\delta \sim 3\times
10^{-5}$ which was finally confirmed by COBE (Bennett et al., 1996) and 
BOOMERanG (Jaffe et al., 2000) on this same level but at larger scales 
(that in turn appeared to be in a successful consistency with another 
famous prediction known as the {\em Harrison-Zel'dovich} scale 
invariant perturbation spectrum). At the same time, the CDP amplitude at Mpc
scale should have been enhanced as indicated by $Ly_{\alpha}$-forest and 
early galaxy formation. This can be achieved by introducing weakly interacting 
massive particles ({\em Cold Dark Matter}): contrary to baryionic damping,
the density contrast of cold collisionless particles grows logarithmically
in the hot Universe boosting the power spectrum at galactic scales
($k > k_{eq}\simeq 0.05\;h\;{\rm Mpc}^{-1}$).

Such an optimistic situation in understanding the structure formation has 
produced a strong impetus for the observational and theoretical cosmology 
extending dramatically by continuing progress in the improved technology of 
deep-sky surveys and CMB temperature detections. The ultimate goal is to 
reconstruct the model parameters and CDP power spectrum from Mpc up to the 
horizon scale, the scope straighforwardly related to the high energy physics 
thus capable of being  observationally tested now.

Three points should be emphasised in connection with the problem of LSS 
formation: theoretical, model, and observational. 

The first point means that LSS formation in the Universe is as 
fundamental problem as the creation of the Universe as a whole: 
both features, the small CDPs and Friedmannian background model  
(the {\em Cosmological Principle}), were produced in a unique 
process of inflation in the very early Universe. The theory works 
at very high energies ($\sim 10^{13}$ GeV) whereas the observations 
occur in a low-energy limit ($\sim 10^{-4}$ eV). To provide a fair 
comparison in such a situation we need a model to know how the
perturbations evolved during the entire history of the Universe. 
Therefore, any confrontation in cosmology between theory and 
observations appears to be model dependent.  

To determine the model we need priorities and parameters. The former 
assumes gravitational instability as the principal mechanism of 
CDP dynamics at large scales, and spatially ramdom Gaussian primordial 
perturbations (both assumptions stay in good agreement with observations). 
The latter assumes knowledge of the current time when the structure is 
observed ($H_0$), the composition and abundance of cosmic matter 
($\Omega_m=\Omega_c +\Omega_\nu +\Omega_b$), dark energy ($\Omega_\Lambda$, 
quintessence, relic scalar field), 
{\em Cosmic Gravitational Waves} (T/S), the number of species of massive 
neutrinos and relativistic particles, the parameters of the perturbation
spectra ($n_S$, $n_T$, $\sigma_8$) and biasing coefficients ($b_{cl}, 
b_{gal}$),
the optical depth at reheating, etc. Nowadays cosmologists venture the 
following approach: if a family of the dark matter models is postulated 
as fairly simple (with a few model parameters) then the recovering of both 
the CDP power spectrum and the cosmological parameters can be provided by 
a given set of observational data. 

Below, we discuss the model under such a conventional probability sense. 
There is no principal restrictions on this way: any theory can be tested 
to the limit if we have enough data. The more data are available the less 
uncertainties remain in the theory and more parameters can be determined.  
Actually, we are now in the beginning of data collection. Cosmologists 
have started the model restoration exercise taking simple theories and 
confronting them with the observational data available. This development 
progresses with an increasing number of model parameters. Theory goes 
from simplicity, however Nature appears complex. 

\section*{DARK MATTER MODELS}

Until recently there were two basic theories claiming to approach 
the corner stones of the LSS formation: inflation and defects. While 
being very much different in their grounds on galaxy seeds -- 
the linear Gaussian scalar perturbations in one case and the non-linear 
non-Gaussian cosmic defects (strings, monopoles, textures) in other case 
-- both models presented the fundamental inevitable perturbations
produced in the very early Universe: the parametrically amplified 
quantum vacuum fluctuations of the inflaton and the topological defects
left after phase transitions in the early Universe, respectively.

However, a simple defect models normalised by the CMB fluctuations have
failed to meet the LSS formation (Watson, 1997). The reason is 
that the non-linear matter perturbations generate all three types of 
the metric fluctuations (scalar, vortex and tensor) which all contribute 
to the Sachs-Wolfe $\Delta T/T$ anisotropy on large angular scale, 
so the resulting S-mode amplitude appears insufficient to develop 
the observed galaxy distribution.

By now  only the inflation theories (adiabatic Gaussian density 
perturbations) with cold dark matter have got through ordeals of fitting 
the LSS and $\Delta T/T$ requirements. The principal quest here is 
the predicted Gaussian nature of small CDPs, which faces a satisfactory 
consistency with the real distribution of galaxies on scales $\sim 20\; 
h^{-1}$ Mpc (e.g. Juszkiewicz and Bouchet, 1996). The only obstacle 
to test reliably this important feature of the CDP seeds is the 
restricted depth of the available galaxy surveys.

Deep galaxy surveys would also be highly welcome for clarifying 
another challenge of the modern cosmology: the fractal model
attacking persistently the cosmological principle. The point is
that large voids seen in the field galaxy and cluster distributions  
extend up to scales $\sim 100\; h^{- 1}$ Mpc which is close to 
catalogues' sizes, thus leaving a room for discussions about the
value of the homogeneity scale (Sylos-Labini et al., 1997). Nevertheless, 
it sould be emphasized that the fractal challenge is rather
an issue for optical galaxies, and not for the distribution of 
the total mass in the Universe. The latter should be pretty 
homogeneous on scales larger than $\sim 10\; h^{-1}$ Mpc to fit beautiful 
Hubble diagrams, to say nothing on the uniform microwave and X-ray 
backgrounds evidencing the cosmic homogeneity at larger scales.

Thus, we consider only models backed by inflationary theories. 
The main tool for Gaussian perturbations is the second moment of 
their spatial distribution related to the power spectrum:

\begin{equation}
\langle \delta^2 \rangle =
\int\limits_0^{\infty}P(k) k^3 dk =
\int\limits_0^{\infty} \Delta_k^2\frac{dk}{k}.
\end{equation}

The dimensionless CDP spectrum $\Delta_k^2$ has a simple meaning of 
the variance of density contrast in the scale $k$ (the wave number) 
within the scale band $dk \sim k$, it is evidently additive ($\delta^2 
\sim \Sigma \Delta_k^2$). 

Before passing to the discussion on the spectrum observational 
reconstruction 
let me sketch briefly the situation with cosmological 
parameters.

\section*{COSMOLOGICAL PARAMETERS}

It seems that the longstanding debate on $H_0=100h\;{\rm km}\;{\rm s}^{-1}
{\rm Mpc}^{-1}$ is approaching to its end and we are going to learn the value 
of the Hubble constant soon. Today, two methods are very promising: 
measuring Cepheids in distant galaxies and the supernovae type Ia. 
The current agreement stands for the value of $h$ between 0.6 and 0.7 
with accuracy $\sim 10\%$, which marginally excludes the matter 
dominated models with a negligible cosmological constant (such models 
would require $h<0.6$) regarding the low limit for the age of the Universe 
coming from globular clusters.

Another issue is the matter abundance in the Universe. 

At the first glance the situation is a following: again we have two groups of 
experiment resulting in different
 conclusions. However, the agreement 
is possible.

The first experiment deals with megaparsec scales -- galaxy halos, groups 
and X-ray clusters. An assumption on the hydrostatic equilibrium within 
cluster cores yields a low dynamical mass responsible for the shape of 
gravitational potential on Mpc scale: $\Omega_m \sim 0.3$. Another 
important observation is a large fraction of baryons inside X-ray 
clusters reaching somehow $\sim 20 \%$ within scale $\sim 1$ Mpc:

\begin{equation}
\frac{M_{b}} {M_m} \sim 0.2, 
\end{equation}
which is also consistent with a low matter density involved dynamically
in small scales: if $\Omega_b \le 0.1$ due to the deuterium measurements, 
$\;\;\Omega_m\sim\Omega_{b}(M_m/M_b){}\,_{\sim}^{<}\, 0.5$ at the dynamical 
scale ($l_D\sim 10\;h^{-1}$ Mpc).

However, some experiments dealing with LSS ($l > l_D$) hints at a matter 
domination ($\Omega_m{}^{>}_{\sim} 0.5$). There are few arguments for it 
(still more model dependent ones in comparison with the small-scale 
arguments):

\begin{itemize}
\item
the existence of  substructures in the majority of galaxy clusters 
evidencing that the clusters are just forming systems, which is possible 
only in the Universe dynamically close to the critical density of matter; 
\item
the large coherence bulk velocities obviously of the cosmological
origin, allowing a principal reconstruction of the total density
contrast (and, in some cases, the consistency with a model 
$\Omega_m \sim 0.5$ and the galaxy 'biasing' factor $b \sim 1$);
\item
the weak gravitational lensing confirming high dynamical mass 
abundance 
around some X-ray clusters;
\item
the lensing argument on the fraction of splitting quasars 
(still much 
dependent on the model assumptions);
\item
the high abundance of $Ly_{\alpha}$ absorption systems 
(the $Ly_{\alpha}$ clouds);
\item
the evolutionary argument on the galaxy clusters number density (still 
under discussion).
\end{itemize}

The low evolution argument about the dependence of cluster abundance 
on 
redshifts,
 which for many years was considered as a basic argument in 
favour 
of a low density Universe, is not any more as strong as it seemed
before. It is the ENACS identification of the nearby galaxy clusters 
(by 
the dispersion velocities of their optical galaxies, see Mazure et al. 
(1996))
 that has shown a previous underestimation of the number density of
nearby Abell clusters. At the moment we may state the consistency of 
the cluster number density evolution curve with a model 
$\Omega_m \sim 0.5$.

A breakthrough in the problem of the model geometry restoration is being 
currently done while building up classical Hubble diagrams (redshifts 
{\em vs} apparent magnitudes) for distant supernovae of type Ia 
(Perlmutter at al., 1998). Contrary to galaxies, such sources look amazingly 
standard candles which is well supported by the distance measurements 
to nearby supernovae. Tested by distant supernovae Ia, the deviations of 
the Hubble diagram from the linear law hint upon the real geometry of 
the Universe. Currently, the predictions are close to a half-to-half 
matter/vacuum Universe ($\Omega_m\sim\Omega_\Lambda\sim 0.5$). However, 
a careful analysis of the systematic effects is required to make this
method really trustable.

While reconstructing cosmological parameters from the CMB temperature 
fluctuations we see the strongest effect coming from the locations and 
amplitudes of the acoustic peaks (the Sakharov oscillations). The results 
of BOOMERanG and MAXIMA-1 have specified the first and second acoustic 
peaks which favours a spatially flat Universe with high abundance of baryons: 
$\Omega_b h^2\simeq 0.03$ (Jaffe et al., 2000). However, better 
understanding of the cosmological model will come after carefull analysis 
of the 2d and 3d acoustic peaks which is a matter of the nearest future. 

Finally, a possible reconciliation between the DM experiments on
small and large scales can be the following: some fraction of dark
matter in the Universe is distributed on large scales and is not 
involved in galaxy halos and groups. 

How can it be arranged?

Today we have purely theoretical ideas on such a possibility.
The most frequently discussed are the models with non-zero 
$\Lambda$-term $(\Omega_\Lambda \in (0.5, 0.7))$, {\em mixed} dark 
matter (cold+hot, with the hot particles like massive neutrinos with a few 
eV rest mass and the corresponding density parameter $\Omega_{\nu} 
\in (0.2, 0.4)$), and a combination of both (the $\Lambda$MDM models). 
In all cases CDM particles form a dynamic structure on Mpc scales, while 
on large scales there is an additional contribution coming from light 
neutrinos or/and vacuum density (the $\Lambda$-term affects 
the cosmological expansion rate and effectively shifts the power spectrum 
of density perturbations towards large scales). 
A sceptical point concerning these and other cosmology models which 
are considered today as possible candidates for the real Universe is 
as follows: all of them are multi-parameter (non-minimal) models; 
the more parameters are involved, 
the better the situation is with the data confrontation. 

Does the latter tell us that we miss something important in our
discussion on the formation of the Universe structure? May be. I
can only conclude here saying than none of the models under
discussion meets all the observational tests. Say, regarding 
previous examples, for $\Lambda \ne 0$ models one can expect
a large fraction of old (relaxed) galaxy clusters and lensed
quasars, whereas the MDM models require $H_0 < 60\; {\rm km}\;{\rm s}^{-1} 
{\rm Mpc }^{-1}$ and too a small abundance of X-ray clusters and 
high-redshift quasars. Not to speak that none of the models considered now
can explain a sharp turn of the spectrum slope at $k\sim 0.05\;h\; 
{\rm Mpc}^{-1}$ indicated by galaxy clusters' distribution in space.
Probably, the dark matter has a more complex nature than we think, existing
in the form of a relic scalar field left after inflation, or varing vacuun 
density, or some other form of interacting relic particles which requires 
special investigation.
In such a situation the observational verifications become extremely 
important. The principal test here is the LSS evolution in the early 
Universe.

\section*{THE SPECTRUM OF DENSITY PERTURBATIONS}

The cosmological models of LSS formation discussed today are aimed to fit 
the observational data at $z = 0$. Thus we cannot distinguish between 
the models without going into their evolution at medium and high redshifts 
where the models demonstrate their essential difference. 

Two main experiments promote a snowball progress in the reconstruction of 
the CDP spectrum, which was impossible in previous years: $\Delta T/T 
(\theta > 1^\prime)$ and direct investigation of the evolution and 
hierarchy of LSSs. The reason for stimulating such a progress is that 
these two experiments confront each other and overlap: the $\Delta T/T$
investigations go nowadays to small comoving scales up to $l \sim 10\;
h^{-1}$ Mpc (recall the corresponding angular scale in arcmin $\theta \sim 
lh$), and, at the same time, we observe a developed structure of clusters, 
filaments, voids, and superclusters reaching the scales $\sim 100\; h^{-1}$ 
Mpc. 

Any reasonable assumption on the "formation" of large voids and
superclusters in Gaussian perturbation theories inevitably leads
to $\Delta T/T$ predictions at $\sim 1^0$ capable of current
detection. It is a puzzle that namely this scale specifies 
the horizon at the decoupling era and therefore the angular scale 
of the first acoustic peak. 

The existence of the acoustic peaks was predicted by theory 
long ago. Now, the time for observations has come: the first peak 
is detected. The time is close when the rest peaks will be found
which allow to precisely formulate and delimit the theory.

We are aware of the cosmological temperature anisotropy on large and
small scales, and have some information on the whole spectrum of the 
CMB fluctuations. 
Fortunately, the small angular scales ($\theta < 1^0$) 
can be effectively tested from the Earth's surface. A hope is that the
terrestrial instruments together with balloon ones as well as the MAP 
and Planck Surveyor satellites, will provide a sensitivity advance 
sufficient for the cosmological model reconstruction.

Meanwhile, the situation with the CDP spectrum  looks rather dramatic. 
At large scales ($\sim 1000\; h^{-1}$ Mpc) the fundamental spectrum 
is small in amplitude and consistent with the HZ slope:

\begin{equation}
\Delta_k^2 \sim k^{3 + n_S},\;\;\;\;n_S = 1.1\pm 0.1.
\end{equation}

However, at smaller scales ($\le 100 \; h^{-1}$ Mpc) the power should be 
boosted as we observe there rich structures in spatial distribution of 
galaxies, clusters, $Ly_{\alpha}$ systems, and distant sources like quasars. 

The latter is especially important. 

We live in the period of the decay of quasar and star formation 
activities (Boyle and Terlevich, 1997). We thus have 
a unique opportunity to observe these numerous early sources tracing
the past dynamics of LSS formation. This would be extremely
informative as the LSS perturbation amplitude, being still less 
than unity today at $l\sim 100\;h^{-1}$ Mpc, was ever lower 
in the past, which would predict a strong inverse evolution of such 
huge systems as superclusters and voids forming due to the
gravitational instability. 

It seems that quasars, the active galactic nuclei of distant
galaxies, form the LSS at medium redshifts ($z \sim 1-2$) which
is provided by their correlation function and the existence of
large QSO groups recalling in properties (the comoving size and
abundance) the local superclusters (Komberg et al., 1996). Actually, 
distant bright quasars may originate in merging galaxies in protoclusters, 
and thus can trace the sites of enhanced matter density at medium and 
high redshifts analogous to how galaxy clusters trace them in the near
space. The dynamical formation of 
these early LSSs suggests that the spectral amplitude at superclusters scale
($\sim 100\; h^{-1}$ Mpc) should be comparable and pretty close
to that at cluster scale ($\sim 10 \; h^{-1}$ Mpc), i.e. the CDP
spectrum is nearly flat between those scales (Komberg and Lukash, 1994):

\begin{equation}
\Delta_k^2 \sim k^{0.9 \pm 0.2}.
\end{equation}
This estimate for the spectrum shape is also backed by the local 
observations of galaxy and galaxy cluster distributions (Guzzo, 1991,
Peacock, 1996, Einasto et al., 1997).

A strong break in the spectrum slope from the HZ asymptotic (3) to
the flat part (4) should have happened at supercluster scale
($\sim 100\; h^{-1}$ Mpc) which is obviously a real feature
of the fundamental CDP spectrum. This 'signature of the God' in 
the primordial spectrum demands its explanation in physics of 
the early Universe.

I cannot help mentioning another connection between the very early Universe 
and the primordial perturbations. This is a possibility to have 
high abundance of cosmic gravitational waves contributing to large-scale 
CMB anisotropy. 

There are at least two reasons for such discussion.

The first one is theoretical. Inflation theory is not discriminative 
to any of the perturbation modes if inflation occurs at GUT energies 
(Lukash and Mikheeva, 1996, 2000): both S (CDP) 
and T (CGW) modes can be 
produced with similar amplitudes and thus 
comparable contribution 
to the CMB anisotropy,

\begin{equation}
\left( \frac{\Delta T}{T}\right)^2_{10^0} = S + T.
\end{equation}

The second reason comes from observations. If the scalar perturbation 
spectrum is 'blue' ($n_S > 1$) then a non-zero T/S$>0$ is required 
to reconcile the COBE $\Delta T/T$ measurement with the galaxy cluster 
abundance (Lukash et al., 2000). 

The problem of T/S is fundamental but can be treated at the moment only 
theoretically. A serious discussion on the observational detection of T/S 
could be launched after CMB polarization measurements, which would require 
the instrumental sensitivity $\sim 1 \;\mu K$ currently marginally reachable.

\section*{CONCLUSIONS AND TENDENCIES}

As never before, cosmologists are close to recovering the real model 
of our Universe and the post-recombination CDP spectrum directly from 
observations, both $\Delta T/T$ and LSS, and to creating an exciting link 
to the physics of the very early Universe. We are going to gain the data 
from the advanced ground and space based CMB explorers as well as large 
surveys of spatial distribution of galaxies, galaxy clusters and quasars,
to delimit the cosmological
 model with unprecedented precision. 

The list of current conclusions may be incomplete:
\begin{itemize}
\item
dark energy is not curvature ($\Omega_\kappa \sim 0$);
\item
quintessence is consistent with $\Lambda$-term;
\item
defects are not responsible for LSS formation;
\item
the slope of the fundamental CDP spectrum is consistent with 
Harrison-Zel'dovich one ($n_S\simeq 1$);
\item
distant supernovae Ia indicate the current acceleration of the
Universe ($\Omega_m<0.5$);
\item
$\Delta T/T$ and LSS data hint on low $\Omega_m{}_\sim^< 0.5$;
\item 
baryonic dark matter is indicated by the $\Delta T/T$ acoustic peaks 
($\Omega_b h^2\simeq 0.03$);
\item
adiabatic $\Lambda$CDM models has survived ($\Omega_m\sim 0.3, 
\;\;\Omega_\Lambda\sim 0.7$);
\item
the best-fit $\Lambda$MDM models predict half-to-half matter/energy
domination ($\Omega_m\sim 0.5, \;\;\Omega_\Lambda\sim 0.5\;$)
(Novosyadlyj et al., 2000).
\end{itemize}

In view of current cosmological discussion we can happily state
a common observational agreement on four important points:
\begin{itemize}
\item
$\Omega_{tot}= 1 \pm 0.1$ (we live in a flat Friedmann Universe);
\item
$n_S= 1 \pm 0.1$ (the spectrum of density perturbations is scale free);
\item
$h = 0.65 \pm 0.1$ (the Hubble constant is close to 
$65\;{\rm km}\;{\rm s}^{-1} {\rm Mpc}^{-1}$);
\item
the observed spectrum of density perturbations is best fitted
by $\Lambda$CDM ($\Omega_m\sim 0.3, \Omega_\Lambda\sim 0.7$) and
$\Lambda$MDM ($\Omega_m\sim 0.5, \Omega_\Lambda\sim 0.5$) models;
\end{itemize}

Nevertheless, the discussion about baryonic and matter abundances
in the Universe is up on the agenda. The critical tests in building
up the true cosmology are the following:
\begin{itemize}
\item
direct detection of the neutrino rest masses and other forms
of dark matter;
\item
determination of the locations and amplitudes of the acoustic peaks in 
$\Delta T/T$;
\item
detection and investigation of the polarization of $\Delta T/T$;
\item
further progress in data analysis of the formation and evolution of
large scale structure in the Universe;
\item
development of the geometrical tests (supernovae Ia, $h$, light elements
abundance).
\end{itemize}

Summarizing, we can state a prominent progress: the detalization and
status of some parts of cosmology remind one an enginearing science level.
Still the fundamental cosmological issues waiting their solutions remain the
following:
\begin{itemize}
\item
the problem of dark matter (multicomponent, baryonic, ...);
\item
the problem of $\Lambda$-term (value, the period of domination, ...);
\item
the break in the CDP spectrum slope at scale $\sim 100\; h^{-1}$ Mpc 
which requires {\em new} physical explanation (scale, physics of dark matter,
why LSS and CMB scales are close to each other ($\sim 100\;h^{-1} Mpc$));
\item
the intensity of cosmic gravity waves (T/S, energy scale of cosmic inflation).
\end{itemize}

\section*{ACKNOWLEDGEMENTS}
The work was supported in part by INTAS (97-1192). The author is grateful
to the Organizing Committee for the hospitality.

\end{document}